\documentclass{elsart}
\usepackage{psfig}

\begin{document}

\begin{frontmatter}
  
\title{$\omega$- and $\phi$-meson production in $pn\rightarrow dV$
    reactions and OZI-rule violation}

\author[ITEP]{L.A. Kondratyuk\thanksref{DFG}},
\author[INR]{Ye.S. Golubeva\thanksref{DFG}} and
\author[Juelich]{M. B\"uscher\thanksref{DFG}}
\address[ITEP]{Institute of Theoretical and Experimental physics, B.\
  Cheremushkinskaya 25, 117259 Moscow, Russia}
\address[INR]{Institute for Nuclear Research, 60th October Anniversary
  Prospect 7A, 117312 Moscow, Russia}
\address[Juelich]{Forschungszentrum J\"ulich, Institut f\"ur
  Kernphysik, 52425 J\"ulich, Germany}
\thanks[DFG]{Supported by DFG and RFFI.}

\begin{abstract}
  
  We investigate the reactions $p n \to d \omega$ and $p n\to d \phi$
  close to threshold and at higher energies. Near threshold we
  calculate the $S$-wave amplitudes within the framework of the
  two-step model which is described by a triangle graph with
  $\pi$-mesons in the intermediate state and find a ratio of the
  $S$-wave amplitudes squared of $R
  =|A_{\phi}|^2/|A_{\omega}|^2=(4-8) \times 10^{-3}$.
  Any significant enhancement of the experimental value of $R(\phi
  /\omega)$ over this prediction can be interpreted as a possible
  contribution of the intrinsic $s\bar{s}$ component in the
  nucleon-wave function.  We present arguments that there is a strong
  resonance effect in the $\omega N$ channel close to threshold.  At
  higher energies we calculate the differential cross sections of the
  reactions $pn\to d\omega$, $pn\to d\phi$ and the ratio of the $\phi
  /\omega$ yields within the framework of the quark-gluon string
  model.  An irregular behavior of the $\phi /\omega$-ratio is found
  at $s \leq 12$ GeV$^2$ due to the interference of the $t$- and
  $u$-channel contributions.


\end{abstract}
\end{frontmatter}
\section{Introduction}
It is well known (see e.g.\ \cite{Lipkin,Ellis1,Ellis2}) that the
ratio of the $\phi /\omega$ yields 
\begin{equation}
R = \frac{\sigma_{\mathrm{A}+\mathrm{B}\to \phi\mathrm{X}}}
         {\sigma_{\mathrm{A}+\mathrm{B}\to \omega\mathrm{X}}}\ ,
\label{eq:phitoomega}
\end{equation}
where the initial and final hadrons do not contain strange quarks, is
a particularly sensitive probe of the OZI rule \cite{Okubo}. At the
standard deviation $\delta = \theta - \theta_{\mathrm{i}} =
0.7^{\circ}$ from the ideal SU(3)$_f$ mixing angle
$\theta_{\mathrm{i}} = 35.3^{\circ}$ we have $R/f = 4.2\times 10^{-3}$
\cite{Ellis2}, where $f$ is the ratio of the phase-space factors.
However, experimental data show an apparent excess of $R/f$ above the
standard value which varies from $(10-30)\times 10^{-3}$ in $\pi N$
and $NN$ collisions to $(100-250)\times 10^{-3}$ in $\bar{N}N$
annihilation at rest and in flight (see e.g.\ the discussion in
\cite{Ellis2}).  In \cite{Ellis2} the big excess of $R$ in $pp$ and
$\bar{p}p$ collisions over the prediction by the OZI rule was treated
in terms of ``shake-out'' and ``rearrangement'' of an intrinsic
$\bar{s}s$ component in the nucleon wave function. On the other hand,
in papers \cite{Locher,Buzatu} the strong violation of the OZI rule in
$\bar{p}p$ annihilation at rest was explained in terms of hadronic
intermediate $K\bar{K}^{\ast}$ states which might create $\phi$
mesons.

Another argument in favour of a large admixture of hidden strangeness
in nucleons was related to an apparently large contribution of the
$\phi$-meson into the isoscalar spectral function which through the
dispersion relation defines the isoscalar nucleon form factor (see
Ref.\cite{Jaffe}). However, as it was shown later (see \cite{Meisner}
and references therein), the main contribution to the isoscalar
spectral function near 1 GeV stems from correlated $\pi \rho$ exchange
which does not involve strange quarks.

Therefore, the question whether there is a large admixture of hidden
strangeness in nucleons seems to be unclarified. Thus, it is important
to investigate such reactions where uncertainties in the
interpretation of $\omega$ and $\phi$ production in terms of
intermediate hadronic states would be comparably small. In this paper
we argue that a good choice in this respect is the reaction
\begin{equation}
pn\to dV\ ,
\label{eq:pn}
\end{equation}
where $V$ denotes the vector mesons $\omega$ and $\phi$.

We show that contributions of the hadronic intermediate states into
the $S$-wave amplitudes of the reactions $p n \to d \phi$ and $p n \to
d\omega$ can be predicted rather reliably using the two-step model
described by triangle graphs with $\pi$ mesons in the intermediate
state \cite{kondrat1,kondrat2}. Therefore, if the $\phi$ and $\omega$
yields will be measured in reaction (\ref{eq:pn}) near threshold
(which e.g.\ can be done at COSY-J\"ulich), the results can be useful
for a better understanding of the OZI-rule violation dynamics.  For
example, any essential deviation from the prediction of the two-step
model could be serious evidence for the above mentioned ``shake out''
or ``rearrangement'' of an intrinsic $\bar{s}s$ component in the
nucleon wave function.

Note that recent measurements of the $\phi /\omega$ ratio in the
reaction $pd\to ^3\!\! HeX$ (performed at SATURNE II
\cite{Wurzinger1,Wurzinger2}) yield
\begin{equation}
  R/f = \left(63 \pm 5\ ^{+27}_{-8}   \right)\times 10^{-3}
  \label{eq:wurzinger}
\end{equation}
which is also clearly above the expectation 4.2$\times 10^{-3}$.
However the dynamics of the reaction $pd\to ^3\!\! HeX$ is not yet
understood. According to \cite{Wilkin1} the two-step model
underestimates the SATURNE data by a factor 2, while according to
\cite{Uzikov} the discrepancy of the two-step model with the data
might be even larger when spin-effects are taken into account.

Experiments on $\omega$ and $\phi$ production in the reaction $ p p
\to p p V$ close to threshold were performed by the SPES3 and DISTO
collaborations at SATURNE (Saclay) \cite{SPES3,DISTO} (see also the
calculations of $\omega$ production in \cite{Speth}).  Near threshold
the dynamics of the reactions $ p p \to p p V$, $ p n \to p n V$ and $
p n \to d V$ are different because the first one is constrained by the
Pauli principle and the two protons in the final state should be in a
$^1 S_0$ state. In the third case the final $pn$ system is in the $^3
S_1$ state while in the second case it can be in both states.
Therefore, a possible violation of the OZI rule is expected to be
different in all those cases.

Another interesting point is that within the framework of the
line-reverse invariance (LRI) assumption the reaction $pn \to dV$ can
be related to the Pontecorvo reaction $\bar{p}d \to VN$.  The data
from the OBELIX and Crystal-Barrel collaborations result in a $\phi /
\omega $ ratio of about $(230 \pm 60)\times 10^{-3}$)
\cite{Sapozhnikov}. Therefore, if LRI is applicable we expect the
violation of the OZI rule in the reaction $pn \to dV$ to be much
larger than it is predicted by the two-step model, which assumes the
dominance of the hadronic intermediate states.

We also consider reaction (\ref{eq:pn}) at higher energies, analyzing
it in terms of Kaidalov's quark-gluon string model \cite{Kaidalov}.
Within the framework of this model the amplitude of reaction
(\ref{eq:pn}) is dominated by three valence-quark exchange in the $t$-
and $u$-channels. This mechanism corresponds to the nucleon Regge-pole
exchange and reaction (\ref{eq:pn}) can be related to other reactions
like $\pi N$ backward scattering and reactions $pp\to d\pi^+$,
$\bar{p}d\to p\pi^-$.  Therefore, the parameters describing the Regge
trajectory and the $t$-dependence of residues can be taken from other
experimental data except for the overall normalization. The latter can
be fixed using the prediction of the two-step model at lower energy.
In this way we can predict the differential cross sections of the
reactions $pn\to d\phi$, $pn\to d\omega$ and $R(\phi /\omega)$ in wide
intervals of energy and momentum transfer.

The paper is organized as follows. In Sect.\ref{sec:twostep} we
analyze the reactions $pn\to d\phi$ and $pn\to d\omega$ near the
thresholds within the framework of the two-step model.  In
Sect.\ref{sec:qgsm} we consider the quark-gluon string model. In
Sect.\ref{sec:conc} we present our conclusions.

\section{The non-relativistic two-step model}
\label{sec:twostep}
\begin{figure}[htb] 
  \begin{center}
    \leavevmode
    \psfig{file=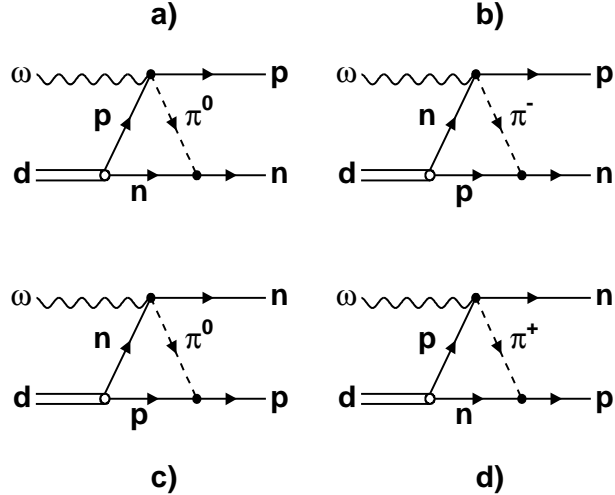,width=8.cm}
    \caption{Diagrams describing the two step model for the case
    $V=\omega$.}  \label{fig:omd} 
  \end{center}
\end{figure}

Let us in the beginning consider the time-reversed reaction
\begin{equation} 
\label{eq:Vd}
Vd\to pn
\end{equation}
where $V$ is the vector meson, $\omega$ or $\phi$. In order to avoid
complications related to the Lorentz-boost of the bound system it is
convenient to calculate the amplitude of reaction (\ref{eq:Vd}) in the
deuteron rest frame. Reaction (\ref{eq:Vd}) is very similar to the
Pontecorvo reaction of meson production in $\bar{p}d$ annihilation
\begin{equation}
\bar{p}d\to MN\ .
\label{eq:pbard}
\end{equation}
The branching ratios for reaction (\ref{eq:pbard}) with $M = \pi$,
$\eta$, $\eta^{\prime}$, $\rho$, $\omega$ and $\phi$ were described in
\cite{kondrat2} within the framework of the two-step model (TSM). The
results are presented in Table \ref{tab:contr}, where it is seen that
the model reproduces BR's for vector meson production rather well.
Here we shall use the TSM for the description of reaction
(\ref{eq:Vd}). In the TSM, which can be described by the diagrams in
Fig.\ref{fig:omd}, reaction (\ref{eq:Vd}) proceeds in two steps: (i)
in the first step the vector meson $V$ creates a $\pi$ meson on one
nucleon through the reaction $\omega N \to N \pi$ and (ii) in the
second step the $\pi$ meson is absorbed by the other nucleon.

First we investigate reaction (\ref{eq:Vd}) at small momenta of the
$\omega$ meson when only the $S$ waves in the $VN$ and $Vd$ systems
contribute significantly. Then the amplitude of the reaction
\begin{equation}
  \label{eq:VN}
  VN\to N\pi
\end{equation}
can be written as 
\begin{eqnarray}
  \langle p_3^{\prime};p_4^{\prime}\lambda_n^{\prime} |
  \hat{T}_{VN\to N\pi} |
  p_1^{\prime},\vec{\epsilon}_{\lambda^{\prime}_{1}}^{\,(V)}
  p_2^{\prime}&& \lambda_2^{\prime} \rangle =  \nonumber \\
  &&\varphi_{\lambda^{\prime}_{4}}^{\ast}(\vec{p}_4^{\,\prime})\ 
  \vec{\epsilon}_{\lambda^{\prime}_{1}}^{\,(V)} \cdot
  \vec{\sigma}\ 
  \varphi_{\lambda^{\prime}_{2}}^{\ast}(\vec{p}_2^{\,\prime})\ 
  A_{VN\to N\pi}(s_1,t_1)\ ,
\label{eq:AVN}
\end{eqnarray} 
where $p_1^{\prime}$, $p_2^{\prime}$, $p_3^{\prime}$ and
$p_4^{\prime}$ are the 4-momenta of the vector meson, the initial
nucleon, the final nucleon and the $\pi$ meson, respectively. The
$\lambda_i^{\prime}$ are the spin projections and the
$\vec{\epsilon}^{\,(V)}$ is the polarization vector of the vector meson
and $s_1=(p_1^{\prime} +
p_2^{\prime})^2=(p_3^{\prime}+p_4^{\prime})^2$, $t_1=(p_1^{\prime} -
p_4^{\prime})^2=(p_2^{\prime}-p_3^{\prime})^2$. The amplitude of
reaction (\ref{eq:Vd}) which corresponds to the triangle graph of
Fig.\ref{fig:omd} has the following form:
\begin{eqnarray}
  \lefteqn{ T_{Vd\to pn}(s,t) =}
  \nonumber \\
  &&\frac{f_{\pi}}{m_{\pi}} \varphi^{\ast}(\vec{p}_4)\ 
  \vec{\sigma}\cdot\vec{\epsilon}^{\,(d)}\ 
  \vec{\sigma}\cdot\vec{\epsilon}^{\,(V)}\ 
  \vec{\sigma}\cdot\vec{M}^{\,(\pi)}\ \varphi(\vec{p}_3)\cdot
  A_{VN\to N\pi}(s_1,t_1) \ ,
\label{eq:AVd}
\end{eqnarray}
where 
\begin{eqnarray}
&M_i^{(\pi)} &= \sqrt{2\,m} \int q_i\,
\Phi_{\pi}(\vec{k},\vec{p}_4)\Psi_d(\vec{k})\, 
\frac{\d^3k}{(2\pi)^{3/2}} \\
&\Phi_{\pi}(\vec{k},\vec{p}) &=
       \frac{F_1(q^2)F_2(q^2)}{q^2-m_{\pi}^2} \label{eq:ff}\\
&q^2 &= m_{\pi}^2 - \delta_0(\vec{k}^2+ 
       \vec{p}\cdot\vec{k} + \beta(\vec{p}_4)))\ , \nonumber \\
&\vec{p} &= -2\,\vec{p}_4/\delta_0\,,\ 
          \vec{q} = \vec{p}_4-\vec{k}\ , \nonumber \\
&\beta(\vec{p}_4) &=
    (\vec{p}_4^{\,2}+m_{\pi}^2-T_4^2)/\delta_0\ , \nonumber \\
&\delta_0 &= 1+T_4/m\ . \nonumber
\end{eqnarray}
Here, $p_1^{\prime}$, $p_2^{\prime}$, $p_3^{\prime}$ and
$p_4^{\prime}$ are the 4-momenta of the vector meson, deuteron, proton
and neutron, respectively, and $s=(p_1+p_2)^2$, $t=(p_1-p_3)^2$,
$u=(p_1-p_4)^2$.  $T_4=E_4-m$ is the proton kinetic energy,
$\Psi_d(\vec{k})$ the deuteron wave function, $\vec{\epsilon}^{\,(d)}$
is the polarization vector of the deuteron, $f_{\pi}\approx 1$ is the
$\pi NN$ coupling constant, $F_1(q^2)$ is the form factor in the $\pi
NN$ vertex, $F_2(q^2)$ is the form factor which cuts virtual masses of
the $\pi$ meson in the amplitude $\omega N\to N\pi$. Note that the
amplitude $M_i^{(\pi)}$ corresponds to the exchange of the neutral
$\pi$ meson (see the left diagrams in Fig.\ref{fig:omd}).

Apart from the $\pi$ exchange some contributions to the $Vd \to pn$
amplitude might also come from other exchanges like $\rho$ or $\omega$
mesons. As it was shown in \cite{kondrat2}, $\rho$ or $\omega$
exchanges give only small contributions to the amplitudes of the
Pontecorvo reactions $\bar{p}d\to MN$.  This can be seen in Table
\ref{tab:contr} taken from \cite{kondrat2}.

\begin{table}[htb]
  \begin{center} 
    \caption{Contributions of different intermediate channels into BR's
      for dif\-fe\-rent Ponte\-corvo reactions ($\times 10^{-6}$)
      within the framework of TSM.  The experimental data are taken
      from \cite{Obelix,Crystal}.}
 \label{tab:contr}
    \begin{tabular}{|c|c|c|c|c|}
      \hline $\bar{p}d \to MN$ & Intermediate & Contribution &
      Contribution& Experiment\\ & state & only $S$-wave & $S$ and $D$
      wave& \\ \hline & $ \pi $ & 12.69 & 12.89 & \\ $\pi^-$p  & $
      \rho $ & 0.12 & 0.09 & \\ & $ \omega$ & 0.02 & 0.01 & \\ & Total
      & 12.8 $\pm$0.4 & 12.9$\pm$ 0.5&$12.9 \pm 0.8$ \\ \hline & $ \pi
      $ & 1.3 & 1.34 & \\ $\eta$n & $ \rho $ & 0.2 & 0.18 & \\ & $
      \omega$ & 0.03 & 0.03 & \\ & Total & 1.6 $\pm$ 0.2 & 1.6 $\pm$
      0.2&$3.19 \pm 0.48$ \\ \hline & $ \pi $ & 2.03 & 2.03& \\ 
      $\eta^{\prime} $n & $ \rho $ & 0.25 & 0.25 & \\ & $ \omega$ &
      0.05 & 0.05 & \\ & Total& 2.3 $\pm$
      0.2 & 2.3 $\pm$ 0.2 &$8.2 \pm 3.4$ \\ \hline & $ \pi $ & 17.86 & 17.91 & \\
      $\rho ^-$p & $ \rho $ & 0.33 & 0.29 & \\ & $ \omega$ & 0.06 &
      0.06 & \\ & Total & 18.3 $\pm$ 2.3 & 18.3 $\pm$ 2.3 &$29.0 \pm
      7.0$ \\ \hline & $ \pi $ & 28.35 & 28.41& \\ $\omega$n & $ \rho
      $ & 0.73 & 0.66 & \\ & $ \omega$ & 0.06 & 0.05 &\\ & Total& 29.1
      $\pm$ 1.7 & 29.1 $\pm$ 1.7 & $ 22.8 \pm 4.1$\\
      \hline & $ \pi $ & 5.54 & 5.54 & \\ $\phi n$ & $ \rho $ & 0.03 & 0.03 &\\
      & $ \omega$ & 0.003 & 0.003 &\\ & Total & 5.6 $\pm$ 0.7 & 5.6
      $\pm$ 0.7 &
      $5.3 \pm 0.9$ \\
      \hline
    \end{tabular} 
  \end{center} 
\end{table} 

The rather strong suppression of those exchanges is related to the
fact that they are mainly contributing at smaller internucleon
distances as compared to $\pi$ exchange. We checked that the
contributions of $\rho$ and $\omega$ exchanges can also be neglected
in the reactions considered here.

It was shown in \cite{kondrat2} that the deuteron $D$-wave
contribution into the amplitudes of Pontecorvo reactions is very small
as compared to the $S$-wave (see Table \ref{tab:contr}). Therefore,
here we take into account only the $S$-wave in the deuteron wave
function. As in \cite{kondrat2} we use the Paris model for the
description of the deuteron wave function, parameterized as in
\cite{Lacomb}.

The product of the two form factors in Eq.(\ref{eq:ff}) we parameterize
using the dipole formula
\begin{equation}
F_1(q^2)F_2(q^2)=\left(\frac{m_{\pi}^2-\Lambda^2}{q^2-\Lambda^2}\right)^2\
.
\end{equation}
The best description of the data on Pontecorvo reactions was found
with $\Lambda = 1.1$~GeV \cite{kondrat2} and here we choose the
cut-off parameter in the interval 1.0 -- 1.2 GeV/c.

At small momenta of the vector meson the differential cross section is
isotropic in the CM system and the amplitude $A_{V}(s_1,t_1)$
can be expressed through the total cross section of the reaction
$\pi^- p \to Vn$ as follows
\begin{equation}
|A_{VN \to N\pi}(s_1,t_1)|^2 = 
\frac{16\,\pi s_1}{3}\
\frac{k_{\pi}^{\mathrm{cm}}(s_1)}{k_V^{\mathrm{cm}}(s_1)}\
\ \sigma(\pi^-p\to Vn)\ .
\label{eq:AV}
\end{equation}
For the cross section $\pi^-p\to n\omega$ we use two different
parameterizations from \cite{Cougnon} and \cite{Sibirtsev} which well
reproduce the available experimental data.  For the cross section
$\pi^-p\to n\phi$ we take the fit from \cite{Sibirtsev}.  Note that
very close to threshold ($k_V^{\mathrm{cm}}(s_1) \leq 100 - 120$
MeV/c) the $S$-wave amplitude of the reaction $\pi^- p \to n\omega$ is
suppressed (see \cite{Binnie}). In the case of the reaction $\pi^- p
\to n\phi$ there is also some indication that
$\sigma/k_V^{\mathrm{cm}}(s_1)$ has an irregular behavior at very
small $k_V^{\mathrm{cm}}(s_1)$. As the origin of this effect is not
completely clear we shall use the TSM for calculation of the cross
sections of the reactions $V d \to pn$ and $p n \to V d$ at those
values of $s$ which correspond to $k_V^{\mathrm{cm}}(s_1)$ larger than
150 MeV/c. The experimental data show that the angular distribution in
the reaction $\pi^- p \to n\omega$ is isotropic and the $S$-wave is
dominant at least until $k_V^{\mathrm{cm}}(s_1) = 260$ MeV/c (see the
comment on p.2805 in \cite{Binnie}). So we shall use the TSM model in the
energy region which corresponds to $k_V^{\mathrm{cm}}(s_1) = 150 -
260$ MeV/c. At other energies we use the predictions of the QGSM,
normalized to the TSM model at an energy corresponding to
$k_V^{\mathrm{cm}}(_1) \simeq 200$ MeV/c.

The differential cross section of reaction (\ref{eq:Vd}) can be
written as
\begin{eqnarray}
\lefteqn{
\frac{\d\sigma_{Vd\to pn}}{\d t} =
}
\nonumber \\
&&
\frac{1}{64\,\pi s}\
\frac{1}{(p_{\omega}^{\mathrm{cm}})^2}\
F(I)\ \overline{|T_{Vd \to pn}(s,t) + T_{Vd\to pn}(s,u) |^2}
\ .
\label{eq:sigmaVd}
\end{eqnarray}
The isospin factor $F(I)$ in Eq.(\ref{eq:sigmaVd}) takes into account the
exchange of $\pi^0$ and $\pi^-$ in the intermediate state and is
defined as 
\begin{eqnarray*}
&|T^{(a)}+T^{(b)}|^2 &= F(I)\,|T^{(a)}|^2\\
&|T^{(c)}+T^{(d)}|^2 &= F(I)\,|T^{(c)}|^2
\end{eqnarray*}
and is equal to 9. The indices (a) to (d) correspond to the four
diagrams in Fig.\ref{fig:omd} and the amplitude $T_{Vd\to pn}(s,u)$ to
the contribution of the lower diagrams in the figure. Near threshold,
where the cross section is isotropic we have
\begin{equation}
\overline{
|T_{Vd\to pn}(s,t) +
 T_{Vd\to pn}(s,u)|^2 } = 
4\,\overline{
|T_{Vd\to pn}(s,t)|^2 }
\end{equation}

The differential cross section of the reaction
\begin{equation}
pn\to dV
\end{equation}
can be expressed through $\d\sigma/\d t$ from Eq.(\ref{eq:sigmaVd}) as
follows
\begin{equation}
\frac{\d\sigma_{pn\to dV}(s,t)}
     {\d t} = 
\frac{9}{4}
\left(
\frac{p_V^{\mathrm{cm}}}{p_p^{\mathrm{cm}}}
\right)^2\
\frac{\d\sigma_{Vd\to pn}(s,t)}
     {\d t}
\label{eq:sigmapn}
\end{equation}

The cross sections of the reactions $pn \to d \phi$ and $pn \to
d\omega$ are rather sensitive to the value of the cut-off parameter
$\Lambda$. We take it in the interval $\Lambda = 1.0$ -- 1.2 GeV.  The
cross section of the reaction $pn\to d\omega$ is rather big and at
$T_p$ = 2.1~GeV ($k_{\omega}^{cm}(s_1) \simeq 220$ MeV/c for the
reaction $\pi^- p \to n\omega$) is in the range 34 -- 60 $\mu$b. The
cross section for $\phi$ production is significantly smaller. It is in
the range 0.16 -- 0.33 $\mu$b at $T_{\mathrm{lab}} = 2.8$ GeV
($k_{\phi}^{cm}(s_1)\simeq 220$ MeV/c for the reaction $\pi^- p \to
n\phi$).  Note that the maximum energy which can be reached at
COSY-J\"ulich is 2.688~GeV.

It is interesting to compare the cross sections of the reactions $p n
\to d \omega$ and $ p p \to p p \omega$ close to threshold.  The last
one was measured by the SPES3 Collaboration at SATURNE (Saclay) at
$T_{\mathrm{lab}}= 1905-1935$ MeV \cite{SPES3} and analyzed within the
framework of the meson-exchange model in Ref.\cite{Speth}.  Adjusting
the cut-off parameter $\Lambda_N$ of the form factor to the data the
authors of Ref.\cite{Speth} calculated the cross section of the
reaction $ p p \to p p \omega$ for proton incident energies up to 2.2
GeV. This model predicts a cross section of about 2 -- 3 $\mu$b at 2.0
GeV.  This is a factor 10 -- 30 smaller than the cross section of the
reaction $p n \to d \omega$ predicted by the two-step model. A similar
situation happens in the case of $\eta$-production near threshold
where the cross section of the reaction $ p n \to d \eta$ is also much
higher than the cross section of the reaction $pp \to pp \eta$ (see
e.g.\ papers \cite{Plouin,Chiavassa,Faldt} and references therein).
This is the consequence of two factors: i) the dominance of the
isovector meson exchange mechanism which gives approximately a factor
6 in favour of the $pn \to pn \eta$ cross section as compared to $pp
\to pp \eta$; ii) the influence of phase space close to threshold
which suppresses the three-body final state as compared with the
two-body state. The value of the suppression depends on the excess
energy $Q = \sqrt{s} - \sqrt{s_0}$, where $\sqrt{s_0}$ is the
CM-threshold energy (see Ref.\cite{Faldt}). At $Q= 20 - 30$ MeV it
might be a factor up to 3 -- 5.  Another factor which enhances the
cross section of the reaction $pn \to dV$ near threshold in our model
is the constructive interference of the amplitudes $A(s,t)$ and
$A(s,u)$.  Very close to threshold where the cross section is
isotropic and the interference of those amplitudes is complete it
gives an enhancement factor of 4.

Let us note that our results for the $pn \to d V$ cross section are
rather close to the prediction of the Rossendorf collision model $\sim
50\ \mu$b at 2.0 GeV \cite{Mueller} as well as to
the estimate made by Wilkin ($\sim 30\ \mu$b at 2.0 GeV
\cite{Wilkin}).

\section{The quark-gluon string model}
\label{sec:qgsm}
\begin{figure}[htb] 
  \begin{center}
    \leavevmode
    \psfig{file=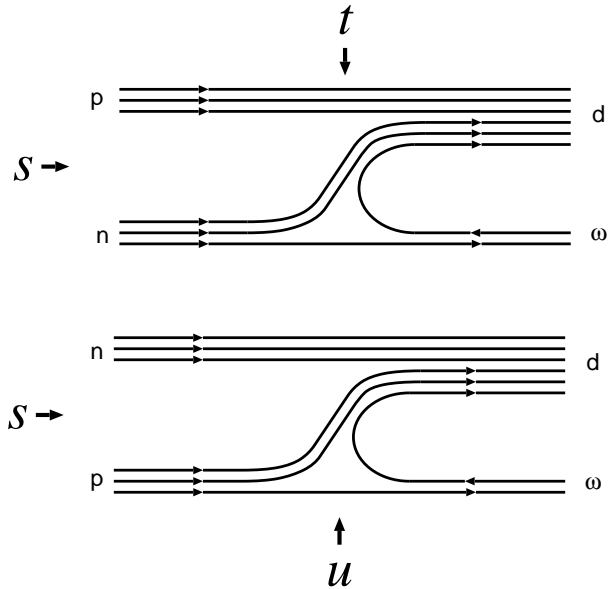,width=8.cm}
    \caption{Diagrams describing three valence quark exchanges in $t$- and
      $u$-channels.}  \label{fig:qgsm}
  \end{center}
\end{figure}

The values of the cross sections presented in the previous section are
only valid near threshold where they are proportional to
$k_V^{\mathrm{cm}}$ and the differential cross sections are isotropic.
In order to describe the angular and energy dependence above threshold
we use the quark-gluon string model (QGSM) developed by Kaidalov in
\cite{Kaidalov}. For a sufficiently high energy $\sqrt{s}$ and not
very high $t$ ($t\leq 1 2\ \mathrm{GeV}^2$) the amplitude of the
reaction $pn\to dV$ is dominated by the exchange of three valence
quarks in the $t$-channel with any number of gluons exchanged between
them (see Fig.\ref{fig:qgsm}). The diagrams of Fig.\ref{fig:qgsm} a)
and b) describe the exchange of three valence quarks in $t$- and
$u$-channels, respectively. The important question is whether we can
apply the QGSM at incident momenta of about 1 GeV/c.

The QGSM is based on two ingredients: i) the topological expansion in
QCD and ii) the space-time picture of the interactions between
hadrons, which takes into account the confinement of quarks.  The
$1/N_c$ expansion in QCD (where $N_c$ is the number of colors) was
proposed by 't Hooft \cite{Hooft}. The behavior of different
quark-gluon graphs according to their topology was analyzed by Rossi
and Veneziano \cite{Rossi}. In the large $N_c$ limit the planar
quark-gluon graphs are dominant. The QGSM proposed by Kaidalov
\cite{Kaidalov} is based on the $1/N_f$ expansion with $N_c \sim N_f$
(where $N_f$ is the number of flavours).  At sufficiently large $N_f$
the simplest planar quark-gluon graphs give dominant contribution into
the amplitudes of binary hadronic reactions. For the amplitudes with
definite quantum numbers in the $t$-channel, the parameter of
expansion is $1/N_f^2 \sim 1/10$. In the space-time representation
those graphs correspond to the formation and break-up of a quark-gluon
string in the intermediate state. The quark-gluon string is identified
with the corresponding Regge trajectory (in case of Fig.\ref{fig:qgsm}
we have in the intermediate state the string with a quark and a
diquark at the ends which corresponds to the nucleon Regge
trajectory). Therefore, the QGSM can be considered as a microscopic
model of the Regge phenomenology and can be used for calculations of
different parameters which were treated before only on
phenomenological level.  In the QGSM each graph can be classified
according to its topology and the amplitude corresponding to this
graph can be considered as analytic function of $s$ and $t$.

As it was shown by Kaidalov \cite{Kaidalov} the QGSM describes rather
well the experimental data on the exclusive and inclusive hadronic
reactions at high energy.  Moreover due to the duality property of
scattering amplitudes \cite{Rossi} this approach can also be applied
in the intermediate energy region (see Ref.\cite{Kaidalov2}),
especially for reactions with no resonances in the direct channel.
In fact this model was successfully applied to the description of the
reactions $pp \to d \pi^+$, $\bar{p}d\to MN$ and $\gamma d \to pn $
at intermediate energies where the diagrams with three valence quark 
exchanges in the $t$-channel were found to be dominant (see
\cite{Kaidalov2,Guaraldo,Desanctis}).  So we expect that the QGSM will
also give a reasonable description of reaction (\ref{eq:pn}).

First of all we have to fix the spin structure of the $pn \to d V$
amplitude. We assume that it is equal to the one predicted by the
two-step model (see Eq.(\ref{eq:AVd})). Then the amplitude
corresponding to the graph of Fig.\ref{fig:qgsm} (upper) can be
written as
\begin{eqnarray}
\lefteqn{
T_{pn\to dV}(s,t) =}
\nonumber \\
&&\varphi^{T \ast}(\vec{n}) (-i \sigma_y)\ 
\vec{\sigma}\cdot\vec{\epsilon}^{\,(d)}\ 
\vec{\sigma}\cdot\vec{\epsilon}^{\,(V)}\ 
\vec{\sigma}\cdot\hat{\vec{p}}\
\varphi(\vec{p})\cdot A(s,t) \ ,
\label{eq:ApndV}
\end{eqnarray}
where $\vec{p}$ and $\vec{n}$ are the proton and neutron momenta
respectively, $\hat{\vec{p}}$ is the unit vector directed along
$\vec{p}$ and $T(s,t)$ is the invariant amplitude, which corresponds
to the nucleon Regge-trajectory
\begin{equation}
  A(s,t)= F(t) \left(\frac{s}{s_0}\right)^{\alpha_{N(t)}} \exp{\left[
      -i\ \frac{\pi}{2}\left(\alpha_{N(t)} -
        \frac{1}{2}\right)\right]}\ .
\label{eq:Tst}
\end{equation} 
Here $\alpha_N(t)$ is the trajectory of the nucleon Regge pole and
$s_0 = m_d^2$. According to the data on $\pi N$ backward scattering
\cite{Lyubimov} it has some nonlinearity:
\begin{equation}
  \alpha_{N(t)}= \alpha_{N(0)} +\alpha'_{N(0)}\, t +
  \frac{1}{2}\, \alpha''_{N(0)}\, t^2 \exp{\left( -\beta t^2 \right)}
  \label{eq:alpha}
\end{equation} 
where $\alpha_{N(0)}=-0.5$, $\alpha'_{N(0)}=0.9\ \mathrm{GeV}^{-2}$
are the intercept and slope of the Regge trajectory, $\alpha''_{N(0)}
= 0.4\ \mathrm{GeV}^{-4}$ and $\beta = 0.03\ \mathrm{GeV}^{-4}$. The
exponential term in (\ref{eq:alpha}) is introduced to prevent the fast
grow of the amplitude with $s$ at large $t$ which would be in
contradiction with unitarity.  The small value of $\beta$ is chosen in
order not to destroy the parameterization of $\alpha(t)$ at $-t \leq
1.6\ \mathrm{GeV}^2$ found in \cite{Lyubimov}.

The dependence of the residue $F(t)$ on $t$ can be taken from
\cite {Kaidalov2,Guaraldo}
\begin{equation}
  F(t) = B {\left[\frac{1}{m^2 - t}\ \exp{(R_1^2t)} + C\, \exp{(R_2^2
        t)} \right]}\ ,
\label{eq:resid}
\end{equation} 
where the first term in the square brackets contains the nucleon pole
and the second term contains also the contribution of nonnucleonic
degrees of freedom in a deuteron.  The parameters of the residue
(except the overall normalization factor $B$) we take the same as in
\cite{Kaidalov2}, which were found by fitting data on the reaction $pp
\to \pi^+ d$ \cite{Allaby} at $-t \leq 1.6\ \mathrm{GeV}^2$:
\begin{equation}
C = 0.7\ \mathrm{GeV}^{-2}, R_1^2 = 3\ \mathrm{GeV}^{-2}\ \mathrm{and}\
R_2^2 = -0.1\  \mathrm{GeV}^{-2}\ .
\end{equation}
Therefore, in our case we have only one free parameter $B$, which we define
normalizing the differential cross section predicted by the QGSM
\begin{eqnarray}
  \lefteqn{ \frac{\d\sigma_{pn\to dV}}{\d t} = }
  \nonumber \\
  && \frac{1}{64\,\pi s}\ \frac{1}{(p_{p}^{\mathrm{cm}})^2}\ 
  \overline{ \ |T_{pn\to dV}(s,t) + T_{pn\to dV}(s,u) |^2 \ }
\label{eq:sigQGSM}
\end{eqnarray}
to the calculated one within the two-step model. As the normalization
point we take $T_p = 2.1$ GeV for $\omega$ production and $2.8$ GeV
for $\phi$ production.

\begin{figure}[htb]     
  \begin{center}
    \leavevmode
    \psfig{file=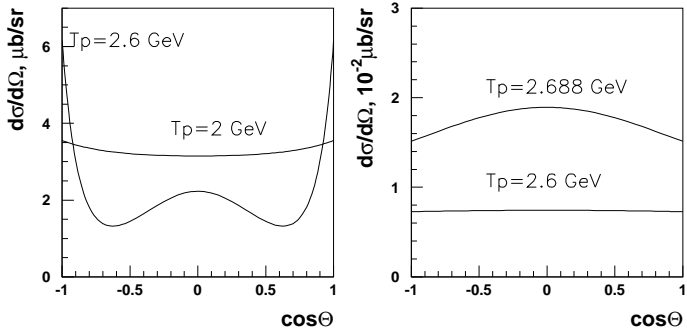,width=9.cm}
    \caption{Differential cross sections of the 
      reactions $pn\to d\omega$ and $pn\to d\phi$.} \label{fig:dsigmado}
  \end{center}
\end{figure}

In Fig.\ref{fig:dsigmado} we present the angular distributions for the
reaction $pn\to d\omega$ at 2.0 and 2.6 GeV and for $pn\to d\phi$ at
2.6 and 2.688 GeV.

We see that the angular distribution for $\omega$ production is almost
isotropic at 2.0 GeV, which demonstrates the dominance of the $S$ wave
at least up to $p_{\omega}^{\mathrm{cm}} = 0.207$ GeV/c.  At 2.6 GeV,
when many partial waves contribute, it has very sharp forward and
backward peaks.  The angular distribution of $\phi$ production is
completely isotropic at 2.6 GeV ($p_{\phi}^{\mathrm{cm}} = 0.07$
GeV/c); it becomes slightly non isotropic at 2.688 GeV
($p_{\phi}^{\mathrm{cm}} = 0.206$ GeV/c) with an excess of about 20\%
at $90^{\circ}$ as compared to the forward and backward angles. It can
be concluded from Fig.\ref{fig:dsigmado} that in both reactions the
$S$ wave dominates up to $p_V^{\mathrm{cm}} \simeq 0.2$ GeV/c.

\begin{figure}[htb] 
  \begin{center}
    \leavevmode
    \psfig{file=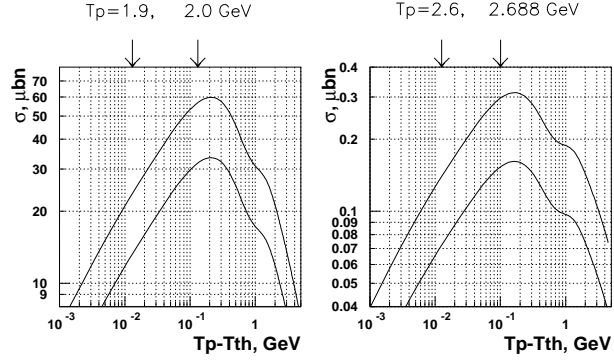,width=8.cm}
    \caption{Energy dependence of the cross sections for the reactions
      $pn\to d\omega$ and $pn\to d\phi$. The upper and
      lower curves are calculated for $\Lambda = 1.2$ and 1.0~GeV,
      respectively.}    \label{fig:sigmapn}
  \end{center}
\end{figure}

The energy dependence of the total cross sections of the reactions $p
n \to d\omega$ and $p n \to d\phi$ is presented in in
Fig.\ref{fig:sigmapn}. The upper and lower curves correspond to
$\Lambda = 1.2$ and 1.0 GeV respectively. Rising above the threshold the
cross sections of both reactions reach the maxima and then decrease
following the power law $(s/s_0)^{2\alpha(0)-2}$. The maximum of the
$\omega$-production cross section is at incident energies covered by
COSY.  The maximum of the cross section for $\phi$ production is very
close to the maximal energy which can be reached at COSY.

\begin{figure}[htb]     
  \begin{center}
    \leavevmode
    \psfig{file=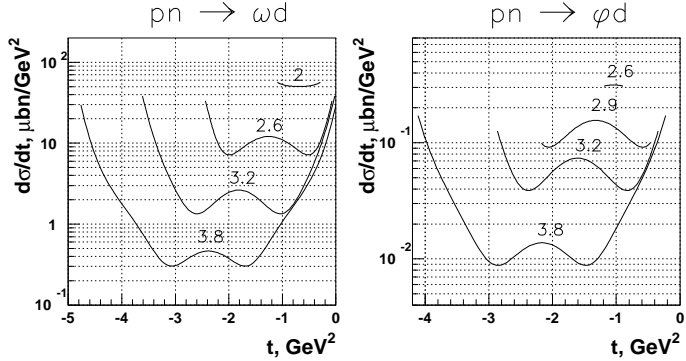,width=9.cm}
    \caption{Differential cross sections of the 
      reactions $pn\to d\omega$ and $pn\to d\phi$ as
      functions of $t$ at energies up to $T_p = 3.8$ GeV.}
      \label{fig:dst}  
  \end{center}
\end{figure}

In Fig.\ref{fig:dst} we present the differential cross sections of the
reactions $pn \to d \omega$ and $pn \to d \phi$ as functions of $t$ at
higher energies up to $T_p$ = 3.8 GeV.  For $\omega$ production the
forward and backward peaks at 2.6, 3.2 and 3.8 GeV can clearly be
seen. For $\phi$ production they are also rather well developed at 3.2
and 3.8 GeV. In the middle of all the curves there is a structure due
to the interference of the diagrams corresponding to the $t$ channel
and $u$ channel exchanges (see Fig.\ref{fig:qgsm}).

Let us now construct the ratio of the differential cross sections
\begin{equation}
  R _{0}(s,t)= \frac{\d\sigma(pn \to d \phi)/\d t}{\d\sigma(pn \to d
  \omega)/\d t}\ .
\end{equation}
In the region where the $t$-channel exchange is dominant and the
$u$-channel exchange can be neglected, the ratio $R_0(s,t)$ would be
independent of $s$ and $t$ in our model. The same will happen in the
region where the $u$-channel exchange is dominant. However, if both
exchanges are significant the value of $R_0(s,t)$ will depend on $s$
and $t$ and this dependence will be strongest at $\theta_{\mathrm{cm}}
= 90^{\circ}$, where the interference is maximal.

\begin{figure}[htb]     
  \begin{center}
    \leavevmode
    \psfig{file=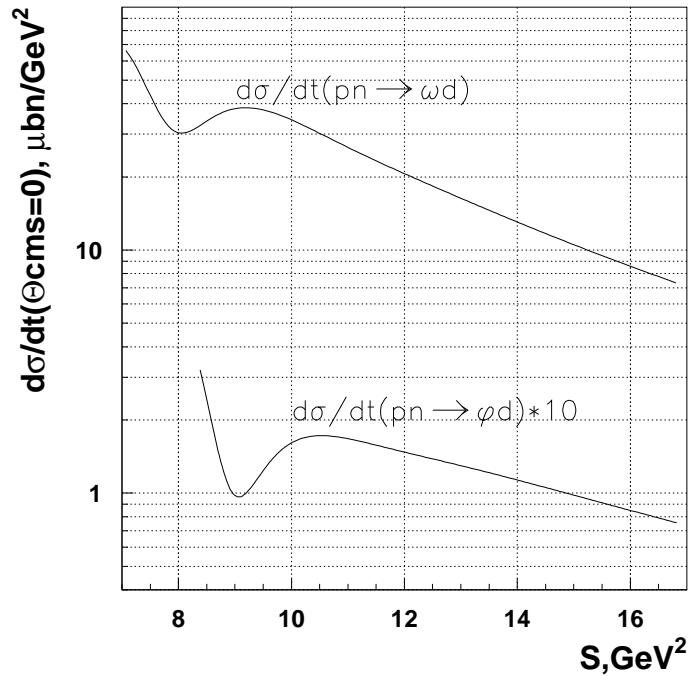,width=9.cm}
    \caption{Differential cross sections
      of the reactions $pn \to d \omega$ and $pn \to d \phi$ at
      $\theta_{\mathrm{cms}}=0^{\circ}$ as functions of $s$.}
    \label{fig:dst0}
  \end{center}
\end{figure}

Certainly, it is more convenient to consider the ratio $R_0(s,t)$ in
those kinematical regions where it is a rather smooth function of the
kinematical variables. From this point of view kinematical regions
where only one mechanism is dominant are preferable.  To illustrate
this point we present in Fig.\ref{fig:dst0} the differential cross
sections of the reactions $pn \to d \omega$ and $pn \to d \phi$ at
$\theta_{\mathrm{cm}}=0^{\circ}$ as functions of $s$.  We see that at
$s \leq 10\ \mathrm{GeV}^2$ the behavior of the differential cross
sections (as well as their ratio) is oscillating which is caused by
the strong interference of the $t$ and $u$ channel exchanges.  However
above $s$ = 12 GeV$^2$ both cross sections decrease with $s$ like
$s^{-3}$ and their ratio $R_0(s,\theta_{\mathrm{cm}}=0^{\circ})$
becomes independent of $s$ and converges to a value of $10\times
10^{-3}$.

\begin{figure}[htb]     
  \begin{center}
    \leavevmode
    \psfig{file=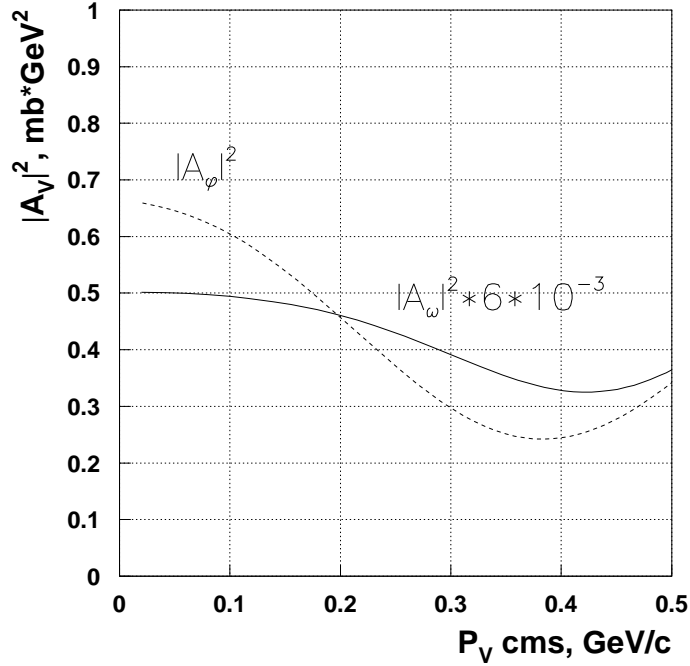,width=9.cm}
    \caption{Invariant amplitudes squared  for the reactions 
      $pn\to d\omega$ and $pn\to d\phi$ as functions of the final
      momentum in the CM system.}  \label{fig:Aphiomeg}
  \end{center}
\end{figure}

Another possibility is to compare the $S$-wave amplitudes of the
reactions $pn \to d \omega$ and $pn \to d \phi$. Near the
corresponding thresholds where the $S$ wave is dominant the $t$ and
$u$ channel exchange amplitudes are equal and do not depend on the
scattering angle.  In Fig.\ref{fig:Aphiomeg} we show the invariant
amplitudes squared $|A_{\omega}|^2$ and $|A_{\phi}|^2$ for both
reactions calculated using our model.  Their ratio is equal
to $6 \times 10^{-3}$ at $p_V^{\mathrm{cm}}= 0.2$ GeV/c. Having in
mind that each cross section has an uncertainty of about 30\%, the
ratio $R$ can be written as:
\begin{equation}
 R =|A_{\phi}|^2/|A_{\omega}|^2=(4-8)\times 10^{-3}.
\end{equation}
Thus the model based on the dominance of hadronic intermediate states
predicts a rather small ratio of the $S$-wave amplitudes for $\phi$
and $\omega$ production.

Another estimate of $R$ can be found if we assume the line-reverse
invariance of the amplitudes, which correspond to the diagrams
presented in Fig.\ref{fig:qgsm}.  In this case we have 

\begin{eqnarray}
  \overline{
    |T^{\mathrm{LRI}}_{pn \rightarrow dV}(s,t)|^2} & = &
    \overline{
      |T^{\mathrm{LRI}}_{pn \rightarrow dV}(s,t) +
      T^{\mathrm{LRI}}_{pn \rightarrow dV}(s,u)|^2} 
    \nonumber \\
    & = &
    \overline{
      |T_{\bar{p}d \rightarrow nV}(s,t) +
      T_{\bar{p}d \rightarrow nV}(s,u)|^2}
\end{eqnarray}    
  
and can define the ratio
\begin{equation}
  R_{\mathrm{LRI}}=|T^{\mathrm{LRI}}_{pn \to d \phi}|^2/
  |T^{\mathrm{LRI}}_{pn \to d \omega}|^2=
  |T_{\bar{p}d \to n \phi}|^2/
  |T_{\bar{p}d \to n \omega}|^2.
\end{equation}

Adopting the result of the OBELIX collaboration $ Y(\bar{p}d \to
n\phi)/ Y(\bar{p}d \to n\omega) = (230 \pm 60)\times 10^{-3}$ we get
\begin{eqnarray}
  R_{\mathrm{LRI}} & = &|T_{\bar{p}d\to n \phi}|^2/|T{\bar{p}d \to n
    \omega}|^2
  \nonumber \\
  & \simeq & (p_{\mathrm{cm}}^{\omega}/p_{\mathrm{cm}}^{\phi})
  (Y(\bar{p}d \to n \phi)/ Y(\bar{p}d\to n\omega)) \simeq (250 \pm
  60)\times 10^{-3},
\end{eqnarray}
which is larger by more than an order of magnitude than that from the
two-step model for the reaction $pn \to dV$.

As the two-step model describes very well the Pontecorvo reactions we
conclude that LRI (and QGSM) is violated in the case of the reactions
$p n \to d\omega$ and $\bar p d \to n\omega$ near threshold. Let us
discuss possible reasons of this violation.

According to \cite{Kaidalov2} the line reverse invariance (LRI) works
fairly well in the case of the reactions $pp \to d\pi^+$ and $\bar{p}d
\to p \pi^-$.  However, there are two regions were the experimental
data on the $pp \to d \pi^+$ total cross section are
noticeably larger than the theoretical expectations:\\
i) at $\sqrt{s} = 2.17$ GeV there is a strong contribution from
intermediate production of the $\Delta$-isobar;\\
ii) around 2.9 GeV there is a resonance-like structure in the cross
section which might be a signal of a broad dibaryon (see e.g.\ 
\cite{kondrat3} and references therein).

Using experimental BR's for the reactions $\bar{p} d \to n\omega$ and
$\bar{p} d \to n\phi$ at rest (see Table \ref{tab:contr}) we can fix
the values of the parameter $B$ in Eq.(\ref{eq:resid}) and calculate
the cross sections of the line-reversed reactions $p n \to d\omega$
and $p n \to d \phi$. In the case of the reaction $p n \to d \omega$
we found $\sigma^{\mathrm{LRI}}_{p n \to d \omega} = 4.3\ \mu$b
at $T_p = 2.0$ GeV, which is about 7 -- 12 times smaller than the
cross section predicted by the two-step model.

This strong violation of LRI in the case of the reactions $p n \to d
\omega$ and $\bar p d \to \omega n$ might indicate that there is a
strong resonance effect in the $\omega N$ channel close to threshold.
A possibility that there might be a resonance in the reaction $\pi^- p
\to \omega n$ near the threshold was discussed earlier in
\cite{Binnie}. It is also known that in the $\rho N$ channel there are
many resonances, which contribute essentially to the cross sections of
the reactions $\pi N \to \rho n$ \cite{PDG} and $\rho N \to \rho N$
\cite{kondrat4} close to threshold.  The baryon resonances with
isospin $I=1/2$ can also couple to the $\omega N$ channel.  

In the case of the reaction $p n \to d \phi$ we have:
$\sigma^{\mathrm{LRI}}_{p n \to d \phi} = 0.22\ \mu$b at $T_p = 2.7$
GeV, which is in perfect agreement with the prediction of the two-step
model.  Therefore, in the case of $\phi$ production LRI is not
violated and we may think that this a hint that resonance effects in
the amplitude $\pi N \to N \phi$ near threshold are absent.  So the
measurements of $\omega$ and $\phi$ production in $pp$, and $pd$
collisions near the thereshold are also interesting from the point of
view of searching for resonances in the $\omega (\phi) N$ channel.

\section{Conclusions}
\label{sec:conc}
Using the two-step model which is described by triangle graphs with
$\pi$-mesons in the intermediate state we calculated the cross
sections of the reactions $pn\to dV$, where $V=\omega$ or $\phi$,
close to threshold and found a ratio of the $S$-wave amplitudes
squared to be $R =|A_{\phi}|^2/|A_{\omega}|^2=(4 - 8)\times 10^{-3}$.
If a significant enhancement of $R(\phi /\omega)$ over the value
predicted by the two-step model will be found in future experimental
data, it might be interpreted as a possible contribution of the
intrinsic $s\bar{s}$ component in the nucleon wave function.
  
The two-step model yields a cross section for the reaction $ p n \to d
\omega$ which is 7 -- 12 times larger than it is predicted with the
help of the LRI assumption from the experimental BR for the reaction
$\bar p d \to \omega n$.  We interpret this result as an evidence of a
strong resonance effect in the $\omega N$ channel close to threshold.
In the case of $\phi$ production near threshold in the reaction $ p n
\to d \phi$ the cross section is in perfect agreement with the
prediction of LRI.

We also analyzed the reaction $p n \to dV$ at higher energies within
the framework of the QGSM, where the amplitude can be described by
contributions of three valence-quark exchange in the $t$ and $u$
channels.  We calculated the differential cross sections of the
reactions $pn\to d\omega$, $pn\to d\phi$ and the ratio of the
$\phi/\omega$ yields in a wide energy interval as functions of $t$.
Essential interference between $t$- and $u$-channel contributions was
found at $s \leq 12\ \mathrm{GeV}^2$, where at small $t$ the $\phi
/\omega$ ratio exhibits irregular behavior as a function of $s$.

\begin{ack}
We are grateful to W.\ Cassing, M.G.\ Sapozhnikov and C.\ Wilkin for
useful discussions.
\end{ack}

\end{document}